# Law for the distance between successive earthquakes

Sumiyoshi Abe

Institute of Physics, University of Tsukuba, Ibaraki 305-8571, Japan

Norikazu Suzuki

College of Science and Technology, Nihon University, Chiba 274-8501, Japan

**Abstract.** Discovery of a new law for the three-dimensional spatial distance between the foci of successive earthquakes is reported. Analyzing the seismic data taken between 1984 and 2001 in southern Californian, it is found that the cumulative distribution of the distances follows the modified Zipf-Mandelbrot law, showing complexity of geometry of the events.

1. **Introduction**

Though seismicity is characterized by extraordinarily complex phenomenology, some of the known empirical laws are remarkably simple. Celebrated examples are the Omori law [*Omori*, 1894] for temporal distribution of aftershocks and the Gutenberg-Richter law [*Gutenberg and Richter*, 1944] for the relationship between frequency and magnitude. These classical examples exhibit the scale-free nature of earthquake phenomena, and this is why seismology has been attracting much attention of researchers of complex systems [*Bak and Tang*, 1989; *Olami et al.*, 1992; *Huang et al.*, 2000; *Bak et al.*, 2002].

In a recent work [*Abe and Suzuki*, 2002], we have studied temporal complexity of earthquakes in southern California. We have analyzed the waiting time distribution between successive earthquakes, and have found that the seismic time series undergoes over a series of transitions between quasi-equilibrium states, each of which obeys the characteristic distribution termed the "$q$-exponential distribution". The $q$-exponential distribution is a generalization of the Zipf-Mandelbrot distribution [*Mandelbrot*, 1983], and is frequently encountered in studies of complex systems at their quasi-equilibrium states. It can be derived, via maximum entropy principle, from Tsallis's generalized entropy [*Tsallis*, 1988; *Tsallis et al.*, 1998; *Abe*, 2002]. The associated generalized statistical theory, referred to as nonextensive statistical mechanics [*Abe and Okamoto*, 2001; *Kaniadakis et al.*, 2002; *Gell-Mann and Tsallis*, in press], is known to be in conformity with the principles of macroscopic thermodynamics, and is currently under active investigation.

Regarding spatial complexity in seismicity, it may be worth pointing out [*Okubo and Aki*, 1987; *Marsan et al.*, 2000] that the distributions of the faults and the foci have the fractal structures. It is of fundamental interest and importance to predict where the focus of the next shock will occur after earlier shocks. Science does not seem to be able to make such a prediction yet but, as a step toward it, it may be worth clarifying the statistical property of the geometric distances between successive earthquakes.

In this paper, we report the discovery of a new empirical law for the nature of spatial geometry of earthquakes. In particular, we analyze the statistical property of the three-dimensional distances between successive earthquakes. We show that the resulting cumulative distribution of the distances is described by the $q$-exponential distributions. This result exhibits spatial complexity of earthquakes in a novel manner.

**2. Modified Zipf-Mandelbrot Law and $q$-Exponential Distribution**

In this section, we discuss the statistical foundations for the Zipf-Mandelbrot law and its modification in view of nonextensive statistical mechanics.

Nonextensive statistical mechanics is a scale-invariant theory, which can explain the statistical properties of a variety of complex systems at their quasi-equilibrium states. It can be formulated by the maximum entropy principle. Previously, in seismology, maximum entropy principle was examined for describing, for example, the frequency-magnitude distribution [*Main and Burton*, 1984], in which the ordinary Boltzmann-Gibbs -Shannon entropy was employed. Here, the entropy functional to be considered is a generalized one termed the Tsallis entropy.

Suppose the three-dimensional distance, $r$, between the foci of successive earthquakes be a fundamental random variable to be measured. $p(r)\,dr$ stands for the probability of finding the value of the distance in the range, $[r, r+dr]$. Therefore, the normalization condition reads

$$\int_0^\infty dr\, p(r) = 1. \qquad (1)$$

In reality, $p(r)$ has a compact support with the finite maximum value, $r_{\max}$.

The Tsallis entropy is given as a functional of $p(r)$ as follows:

$$S_q[p] = \frac{1}{1-q}\left(\int_0^\infty \frac{dr}{\sigma}[\sigma p(r)]^q - 1\right), \qquad (2)$$

where $q$ and $\sigma$ are the positive entropic index and a scale factor of the dimension of length, respectively. In the limit $q \to 1$, this quantity converges to the Boltzmann-Gibbs-Shannon entropy:

$$\lim_{q \to 1} S_q[p] = -\int_0^\infty dr\, p(r)\ln[\sigma p(r)]. \qquad (3)$$

Henceforth, $\sigma$ is set equal to unity for simplicity. The Tsallis entropy is known to share a lot of common properties with the Boltzmann-Gibbs-Shannon entropy, including concavity and fulfillment of the *H*-theorem. However, additivity is violated. For a system composed of two statistically independent systems, *A* and *B*, the Tsallis entropy satisfies

$$S_q(A, B) = S_q(A) + S_q(B) + (1-q) S_q(A) S_q(B), \tag{4}$$

which is referred to as pseudoadditivity. The last term on the right-hand side brings the origin of nonextensivity of the resulting generalized statistical mechanics. Therefore, the value, $|1-q|$, indicates the degree of nonextensivity in a complex system under consideration.

In nonextensive statistical mechanics [*Tsallis et al.*, 1998; *Abe and Okamoto*, 2001], the definition of the expectation value should also be generalized to the "*q*-expectation value", according to the generalization of the entropy. The expectation value of the distance is given by

$$<r>_q = \int_0^\infty dr\, r\, P_q(r), \tag{5}$$

where $P_q(r)$ is the escort distribution [*Beck and Schlögl*, 1993] defined by

$$P_q(r) = \frac{[p(r)]^q}{\int_0^\infty dr' \, [p(r')]^q}. \tag{6}$$

Now, maximization of the Tsallis entropy under the constraints on the normalization in Eq. (1) and the $q$-expectation value of the distance in Eq. (5) leads to the following optimal distribution [*Tsallis et al.*, 1998; *Abe and Okamoto*, 2001]:

$$p_0(r) = \frac{1}{Z_q} e_q\left(-\frac{\beta}{c}(r - <r>_q)\right), \tag{7}$$

$$Z_q = \int_0^\infty dr \, e_q\left(-\frac{\beta}{c}(r - <r>_q)\right), \tag{8}$$

$$c = \int_0^\infty dr \, [p_0(r)]^q. \tag{9}$$

Here, $\beta$ is the Lagrange multiplier associated with the constraint on the $q$-expectation value of the distance. $e_q(x)$ denotes the "$q$-exponential function" defined by

$$e_q(x) = \begin{cases} [1 + (1-q)x]^{1/(1-q)} & (1 + (1-q)x \geq 0) \\ 0 & (1 + (1-q)x < 0) \end{cases}, \tag{10}$$

whose inverse is the "$q$-logarithmic function"

$$\ln_q(x) = \frac{1}{1-q}\left(x^{1-q} - 1\right). \tag{11}$$

In the limit $q \to 1$, the $q$-exponential and $q$-logarithmic functions tend to the ordinary exponential and logarithmic functions, respectively.

$p_0(r)$ in Eq. (7) is a power-law distribution unless in the limit $q \to 1$. It has a long tail if $q > 1$, whereas there appears the cut-off at $r_c = r_0/(1-q)$ if $0 < q < 1$, where $r_0 = [c + (1-q)\beta <r>_q]/\beta$, which is positive even if $q > 1$ [*Abe and Rajagopal*, 2000a]. Therefore, $r_{max} = r_c$ if $0 < q < 1$, whereas $r_{max}$ can be arbitrarily large if $q > 1$. Using this $r_0$, $p_0(r)$ is rewritten as follows:

$$p_0(r) = \frac{e_q(-r/r_0)}{\int_0^\infty dr'\, e_q(-r'/r_0)}. \tag{12}$$

This is referred to as the "$q$-exponential distribution". It is a generalization of the Zipf-Mandelbrot distribution. (The standard Zipf-Mandelbrot distribution corresponds to the case when $q > 1$.) The exponential distribution is seen to be recovered in the limit $q \to 1$.

A point of crucial importance in nonextensive statistical mechanics is that the quantity to be compared with the observed distribution is not $p_0(r)$ but its associated escort distribution [*Abe and Rajagopal*, 2000b]. Accordingly, the cumulative distribution should be defined by

$$P(>r) = \int_r^\infty dr' \, P_q(r').  \qquad (13)$$

Using Eqs. (6) and (12), we obtain

$$P(>r) = e_q(-r/r_0).  \qquad (14)$$

In what follows, we shall show by data analysis that the statistical property of the three-dimensional distances between the foci of successive earthquakes is described extremely well by the distribution in Eq. (14) with $0 < q < 1$, exhibiting spatial complexity of earthquakes.

## 3. Distribution in Southern California

In this section, we present the data analysis and its comparison with the modified Zipf-Mandelbrot law.

We have analyzed the earthquake catalog made available by the Southern California Earthquake Data Center (http://www.scecdc.scec.org/catalogs.html) covering the period between 00:25:8.58 on 1 January 1984 and 23:44:2.81 on 31 December 2001 in the region spanning $29°15.25'\,\text{N} - 38°49.02'\,\text{N}$ latitude and $113°09.00'\,\text{W} - 122°23.55'\,\text{W}$ longitude. (We have taken this period since the data in 1983 are partially missing for a few months.) The number of events is 364867. The data contain, not only significant earthquakes but also extremely weak earthquakes like magnitude 0.0. We have calculated the three-dimensional distances between successive earthquakes.

In Figures 1a and 1b, we present the log-log and semi-log plots of the cumulative distribution associated with the statistical frequency of the distance, respectively. The dots represent the observed distribution, whereas the solid line corresponds to the Zipf-Mandelbrot law described by the $q$-exponential function. We also present the semi-$q$-log plot in Figure 1c. It is clearly appreciated that quality of the fitting is extremely high.

## 4. Distribution in Japan

For comparison with the result obtained in the previous section, we here present the analysis of the catalog of earthquakes in Japan made available by the Japan University Network Earthquake Catalog (http://kea.eri.u-tokyo.ac.jp/CATALOG/junec/monthly.html) covering the period between 01:14:57.63 on 1 January 1993 and 20:54:38.95 on 31 December 1998 in the region spanning $25.851°\,\text{N} - 47.831°\,\text{N}$ latitude and $126.433°\,\text{E} - 148.000°\,\text{E}$ longitude. The number of events is 123390. (We have limited ourselves to this period since before 1993 the number of the observed data per year turned out to be about half of the latter period. An essential difference of the catalog of the Japan University Network Earthquake Catalog from that of the Southern California Earthquake Data Center is that the former does not contain earthquakes with magnitudes smaller than 2, unfortunately.)

In Figures 2a and 2b, we present the log-log and semi-log plots of the cumulative distribution associated with the statistical frequency of the distance, respectively. The dots represent the observed distribution, whereas the solid line corresponds to the Zipf-Mandelbrot law described by the $q$-exponential function. We also present the semi-$q$-log plot in Figure 2c. As in the case in southern California, the fitting is seen to be very well.

## 5. Conclusions

We have studied the statistical property of the three-dimensional distances between successive earthquakes and have discovered that it obeys the modified Zipf-Mandelbrot law characterized by the $q$-exponential distributions with $q = q_s$ less than unity ($q_s = 0.75 \sim 0.77$). This result exhibits complex spatial geometry of earthquake phenomenon in a novel manner.

In our previous work on the time intervals between successive earthquakes [*Abe and Suzuki*, 2002], the associated waiting time distribution was shown to be given also by the $q$-exponential distribution with $q = q_t > 1$ (typically, $q_t = 1.2 \sim 1.3$). It is of interest to notice that the *duality relation*, $q_s + q_t \sim 2$, might hold, though it is still hypothetical.

**Acknowledgments.** We would like to thank Professor G. Igarashi for informative discussions. S. A. was supported in part by the internal research project of the University of Tsukuba.

**References**

Abe, S., Stability of Tsallis entropy and instabilities of Rényi and normalized Tsallis entropies: A basis for $q$-exponential distributions, *Phys. Rev. E*, *66, 046134*, 2002.

Abe, S., and Y. Okamoto, eds., *Nonextensive Statistical Mechanics and Its Applications*, Springer-Verlag, Heidelberg, 2001.

Abe, S., and A. K. Rajagopal, Rates of convergence of non-extensive statistical distributions to Lévy distributions in full and half-spaces, *J. Phys. A*, *33*, 8723-8732, 2000a.

Abe, S., and A. K. Rajagopal, Microcanonical foundation for systems with power-law distributions, *J. Phys. A*, *33*, 8733-8738, 2000b.

Abe, S., and N. Suzuki, Zipf-Mandelbrot law for time intervals of earthquakes, e-print 2002 available at http://xxx.lanl.gov/abs/cond-mat/0208344

Bak, P., K. Christensen, L. Danon, and T. Scanlon, Unified scaling law for earthquakes, *Phys. Rev. Lett.*, *88*, 178501, 2002.

Bak, P., and C. Tang, Earthquakes as a self-organized critical phenomenon, *J. Geophys. Res.*, *94*, 15635-15637, 1989.

Beck, C., and F. Schlögl, *Thermodynamics of Chaotic Systems: An Introduction*, Cambridge University Press, Cambridge, 1993.

Gell-Mann, M., and C. Tsallis, eds., *Nonextensive Entropy \Interdisciplinary Applications*, in press Oxford university Press, Oxford, 2003.

Gutenberg, B., and C. F. Richter, *Bull. Seism. Soc. Am.*, *34*, 185-188, 1944.

Huang, Y., A. Johansen, M. W. Lee, H. Saleur, and D. Sornette, Artifactual log-periodicity in finite size data: Relevance for earthquake aftershocks, *J. Geophys. Res.*, *105*, 25451-25471, 2000.

Kaniadakis, G., M. Lissia, and A. Rapisarda, eds., special issue of *Physica A*, *305*, 2002.

Main, I. G., and P. W. Burton, Information theory and the earthquake frequency-magnitude distribution, *Bull. Seism. Am.*, *74*, 1409-1426, 1984.

Mandelbrot, B. B., *The Fractal Geometry of Nature*, Freeman, San Francisco, 1983.

Marsan, D., C. J. Bean, S. Steacy, and J. McCloskey, Observation of diffusion processes in earthquake populations and implications for the predictability of seismicity systems, *J. Geophys. Res.*, *105*, 28081-28094, 2000.

Okubo, P. G., and K. Aki, Fractal geometry in the San Andreas fault system, *J. Geophys. Res.*, *92*, 345-355, 1987.

Olami, Z., H. J. S. Feder, and K. Christensen, Self-organized criticality in a continuous, nonconservative cellular automaton modeling earthquakes, *Phys. Rev. Lett.*, *68*, 1244-1247, 1992.

Omori, F., On the aftershocks of earthquakes, *J. Coll. Sci. Imp. Univ. Tokyo*, *7*, 111-216, 1894.

Tsallis, C., Possible generalization of Boltzmann-Gibbs statistics, *J. Stat. Phys.*, *52*, 479-487 (1988).

Tsallis, C., R. S. Mendes, and A. R. Plastino, The role of constraints within generalized nonextensive statistics, *Physics A*, *261*, 534-554 (1998).

# Figure Captions

Figure 1a  Log-log plot of the cumulative distribution of the three-dimensional distances between successive earthquakes in southern California. The values of the index, $q$, and the parameter, $r_0$, are $q = 0.773$ and $r_0 = 1.79 \times 10^2$ km, respectively.

Figure 1b  Semi-log plot of the data in Figure 1a.

Figure 1c  Plot of the data in Figure 1a on semi-$q$-log scale. The straight line describes the $q$-exponential distribution. The value of correlation coefficient is $\rho = -0.9993$.

Figure 2a  Log-log plot of the cumulative distribution of the three-dimensional distances between successive earthquakes in Japan. The values of the index, $q$, and the parameter, $r_0$, are $q = 0.747$, $r_0 = 5.95 \times 10^2$ km, respectively.

Figure 2b  Semi-log plot of the data in Figure 2a.

Figure 2c  Plot of the data in Figure 2a on semi-$q$-log scale. The straight line describes the $q$-exponential distribution. The value of correlation coefficient is $\rho = -0.9990$.

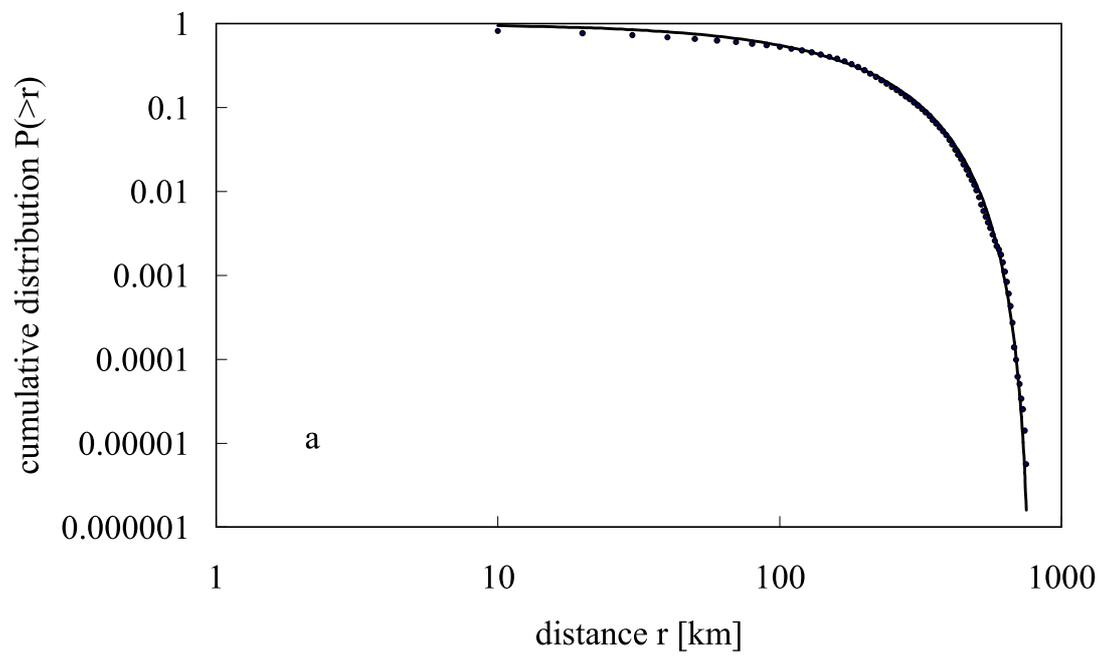

Figure 1a

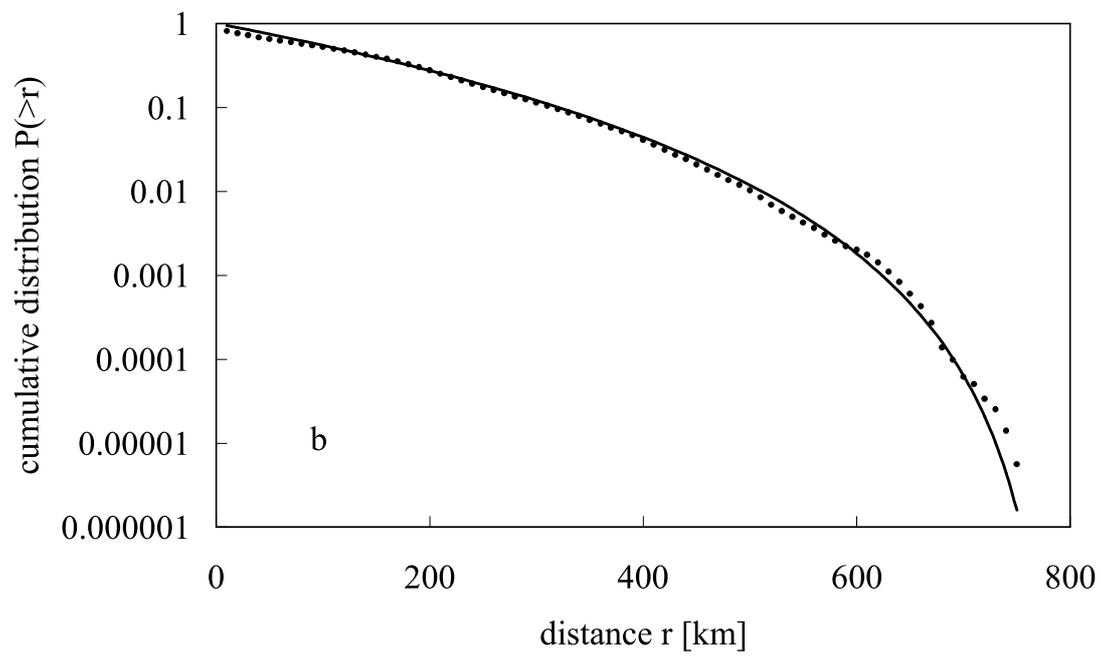

Figure 1b

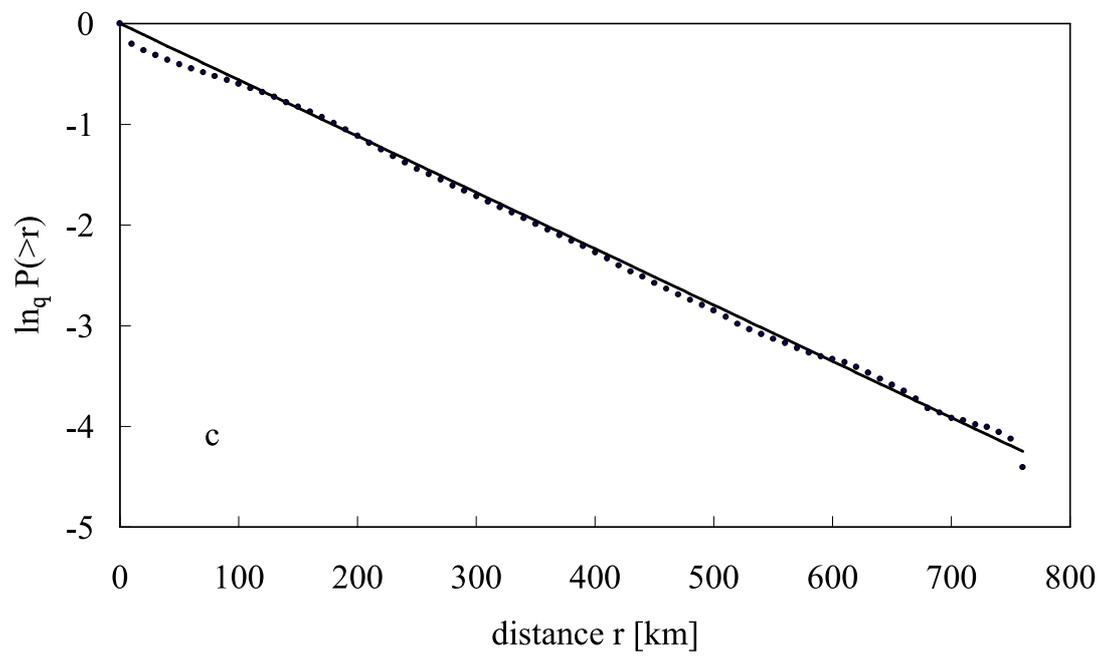

Figure 1c

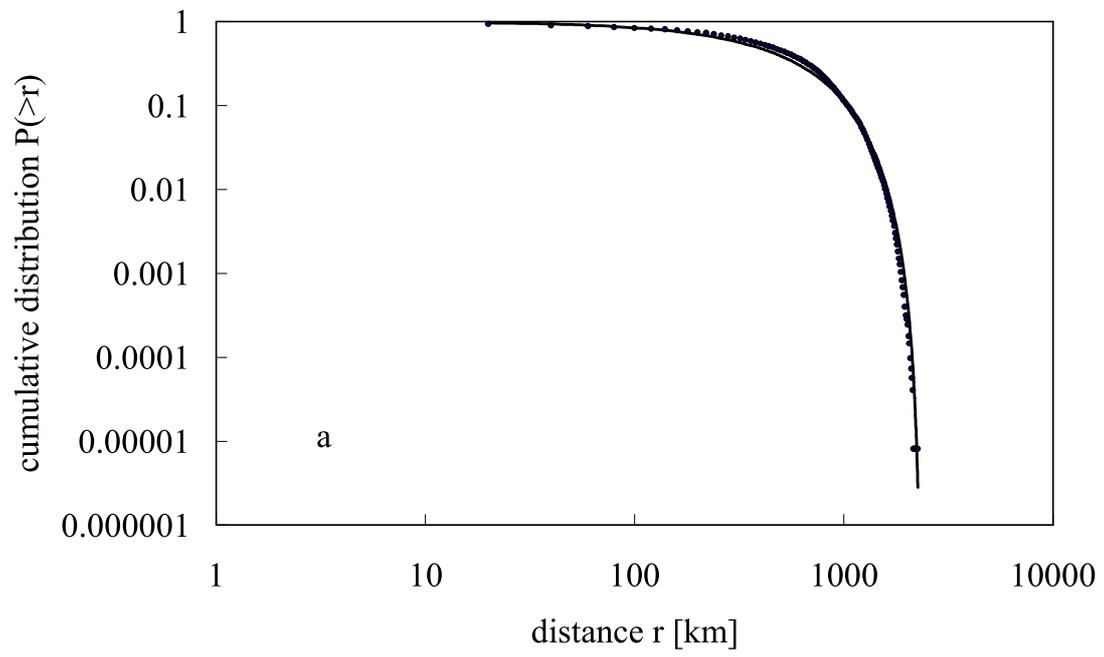

Figure 2a

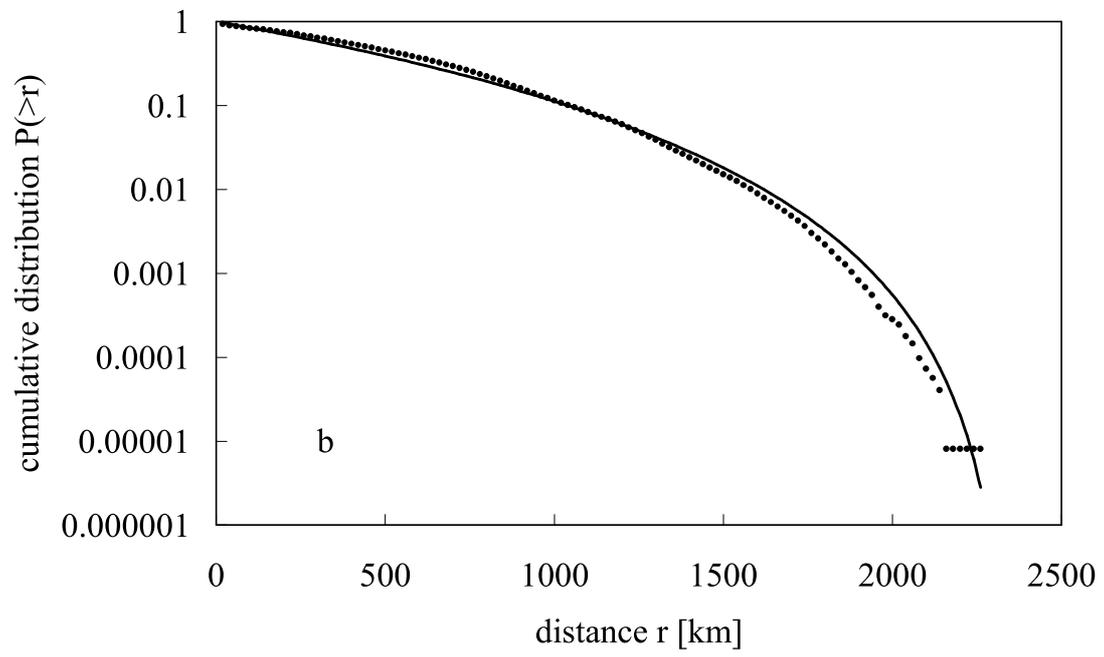

Figure 2b

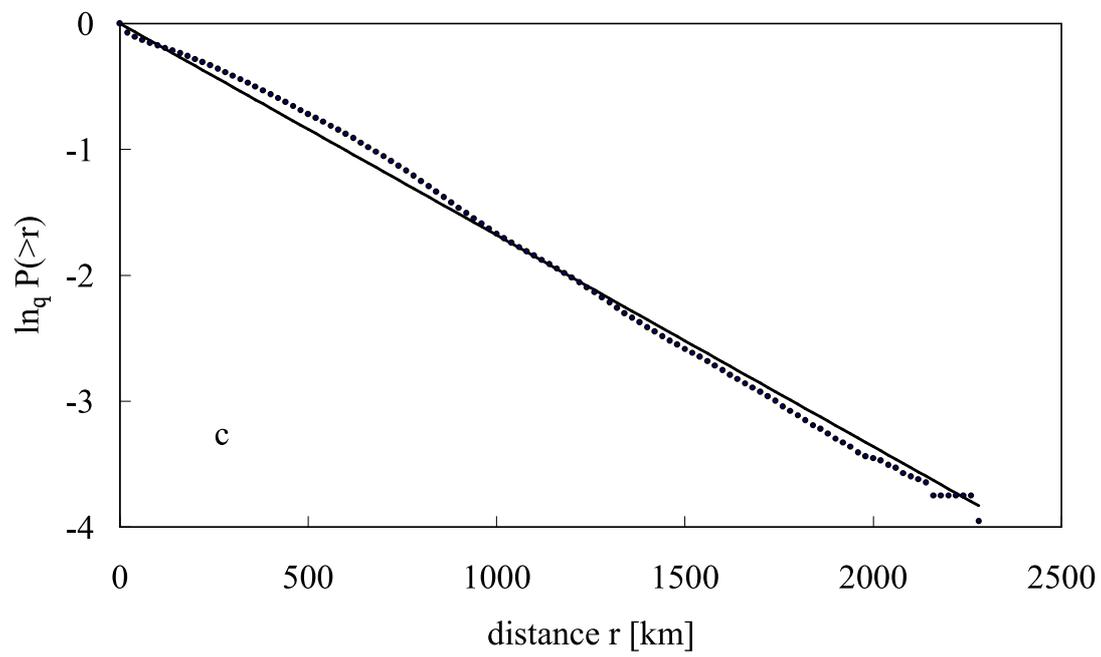

Figure 2c